\documentstyle[twoside,fleqn,espcrc2]{article}

\newcommand{\AmS}{{\protect\the\textfont2
  A\kern-.1667em\lower.5ex\hbox{M}\kern-.125emS}}

\title{Rapidity spectra analysis in terms of non-extensive statistic
approach}

\author{M.Rybczy\'nski, Z.W\l odarczyk
        \address{Institute of Physics, \'Swi\c{e}tokrzyska Academy,
        Kielce, Poland \\
        emails: mryb@pu.kielce.pl and wlod@pu.kielce.pl}
        and
        G. Wilk\address{The Andrzej Soltan Institute for Nuclear Studies,
        Nuclear Theory Department, Warsaw, Poland\\
        email: wilk@fuw.edu.pl}}

\begin{document}

\begin{abstract}
We provide description of rapidity spectra of particles produced in
$p\bar{p}$ collisions using anomalous diffusion approach to account
for their non-equilibrium character. In particular, we exhibit
connection between multiproduction processes and anomalous diffusion
described through the nonlinear Focker-Planck equation with
nonlinearity given by the nonextensivity parameter $q$ describing the
underlying Tsallis $q$-statistics and demonstrate how it leads to the
Feynman scaling violation in these collisions. The $q$ parameter
obtained this way turns out to be closely connected to parameter
$1/k$ converting the original poissonian multiplicity distribution to
its observed Negative Binomial form. The inelasticity of reaction has
been also calculated and found to slightly decrease with the
increasing energy of reaction.
\end{abstract}

\maketitle

\section{INTRODUCTION}

In description of multiparticle production processes one often uses
statistical methods and concepts which follow the classical
Boltzmann-Gibbs (BG) approach. However, it was demonstrated recently
that to account for the long range correlations and for some
intrinsic fluctuations in the hadronizing system one should rather
use the nonextensive Tsallis statistics \cite{T}, in which one new
parameter $q$ describes summarily the possible departure from the
usual BG case (which is recovered in the $q\rightarrow 1$ limit)
\cite{WWq}. Here we shall provide detail description of the rapidity
spectra of particles (mostly pions) produced in $p\bar{p}$ collisions
using the anomalous diffusion approach to this problem and in this
way accounting for their non-equilibrium character.

This method originates from diffusion model approach to nuclear
multiparticle production collisions developed in \cite{Wol}, which
has been further (successfully) applied to the recent RHIC data in
\cite{Biya}. The first attempt to extend it also to the case of
anomalous diffusion (which corresponds to nonextensive $q\neq 1$
case) has been presented recently for nuclear collisions \cite{Lav}.
The non-linear Fokker-Planck (FP) equation used in this case has form
($f=f(y,t)$ with $y$ being rapidity and $t$ time variable):
\begin{equation}
\frac{\delta}{\delta t}f^{\mu} =
\frac{\delta}{\delta y}\left\{ J(y)f^{\mu}
       + D \frac{\delta}{\delta y}f^{\nu}\right\},
\end{equation}
$D$ and $J$ are diffusion and drift coefficients respectively.
The hadronization process is visualized here as diffusion in the
rapidity space starting with rapidities (in cms frame)
$Y^{(\pm)}_{max} \simeq \pm \ln\frac{\sqrt{s}}{m_T}$ (where
$\sqrt{s}$ is invariant energy of reaction and $m_T=\sqrt{m^2 +
\langle p_T\rangle^2}$ mean transverse mass kept here as given. In
the approaches presented in \cite{Wol} and \cite{Biya} linear or
constant drift coefficients has been used and in effect obtained
double gaussian-like form of rapidity spectra\footnote{Interesingly
enough, such spectra for $\frac{dN}{dy}$ were already postulated and
used on purely phenomenological grounds as simple parametrizations of
results of string models allowing for fast numerical calculations of
cosmic ray cascades \cite{KlarCap}.}. As was demonstrated in
\cite{Buk,Lav} for drift proportional to the longitudinal momentum of
the particle (i.e., $J(y) \sim \sinh y$) one gets the thermal
(Boltzmann) distribution (for linear FP equation, i.e., for
$q=\nu=1$) whereas for nonlinear-FP equation the time dependent
solution has the specific power-like form the norm of which is
conserved only for $\mu=1$, therefore $\nu=2-q$ and our distribution
is given by the following formula:
\begin{equation}
f_{q}(y) = \left[ 1-(1-q)\frac{m_{T}}{T} \cosh\left( y-y_{m} \right)
\right]^{\frac{1}{1-q}}. \label{eq:form}
\end{equation}

\section{RESULTS}
We show now results of fitting experimental data on rapidity
distributions of secondaries produced in $p\bar{p}$ collisions
\cite{Gim} by using formula (\ref{eq:form}) in the following form:
\vspace{-3cm}
\begin{figure}[h]
\setlength{\unitlength}{1cm}
\begin{picture}(8.0,8.0)
\includegraphics{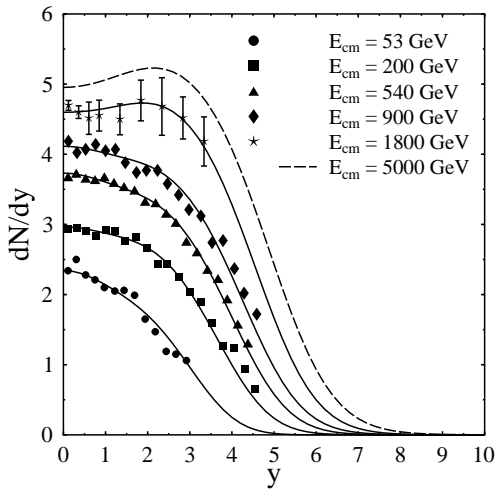}
\end{picture}
\caption{Rapidity spectra obtained by UA5 \cite{UA5} and Tevatron
\cite{Tevatron} fitted by formula (\ref{eq:Fq}). The broadest
spectrum (full line without points) shows our extrapolation to the
LHC energy range (5 TeV).}
\end{figure}
\vspace{-0.7cm}
\begin{equation}
f_{q}(y) = f^{(+)}_q(y) + f^{(-)}_q(y) \label{eq:Fq}
\end{equation}
where
\begin{equation}
f^{(\pm)}_q(y) = \left[ 1-(1-q)\frac{m_{T}}{T}\cosh\left(y\pm
y_{m}\right)\right]^{\frac{1}{(1-q)}}
\end{equation}
with $T$, $y_m$ and $q$ being the (energy dependent) parameters
(the values of transverse mass $m_T$ was in this case taken as fixed
and equal to $m_T = 0.4$ GeV, its possible energy dependence (for
example such as used in (\ref{eq:inel}) below) is then hidden in the
energy dependence of the parameter $T$). The results are shown in
Fig. 1 (for UA5 \cite{UA5} and Tevatron \cite{Tevatron} data) where
the following energy dependences of $T$, $y_{m}$ and $q$ were used:
\begin{eqnarray}
T(\sqrt{s}) &=& a_1 + b_1\cdot\ln \left( \sqrt{s} \right), \nonumber\\
y_{m}(\sqrt{s}) &=& a_2+b_2\cdot \ln\left(\sqrt{s}\right),
\label{eq:param}\\
q(\sqrt{s}) &=& a_3 + b_3\cdot \ln\left(\sqrt{s}\right).\nonumber
\end{eqnarray}
The values of coefficients are: $a_{1}=-0.178\pm0.099$,
$b_{1}=0.267\pm0.016$, $a_{2}=0.451\pm0.129$,
$b_{2}=0.239\pm0.023$, $a_{3}=0.91\pm0.012$, $b_{3}=0.072\pm0.002$.
It turns out that the same parameters can also fit data from P238
\cite{P238} and UA7 \cite{UA7} groups obtained for different rapidity
regions at energy $630$ GeV, cf. Fig. 2. It is interesting that
although our $y_m$ is not supposed to be {\it a priori} connected
with the values of the maximal possible rapidities at given energy,
$Y_{max}$, nevertheless we have found that ratio $y_m/Y_{max}$
changes only weakly with energy varying between $53$ and $1800$ GeV
from $0.265$ to $0.285$. Notice also that for small values of
$|y-y_m|$ Eq.(\ref{eq:Fq}) can be aproximated by double
quasi-gaussian form (which becomes pure double gausian function in
the limit of $q\rightarrow 1$). \\
\vspace{-3.3cm}
\begin{figure}[h]
\setlength{\unitlength}{1cm}
\begin{picture}(8.0,8.0)
\includegraphics{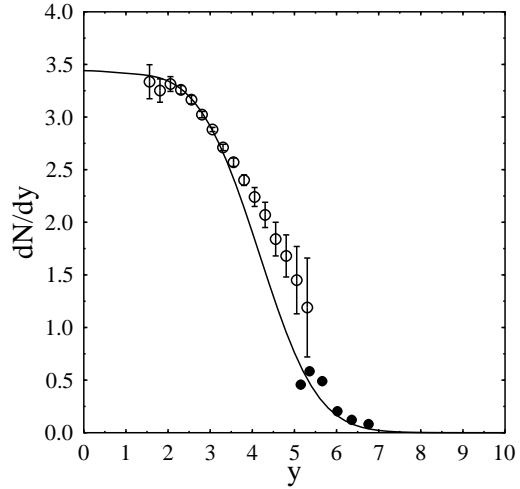}
\end{picture}
\caption{Comparison of data for rapidity distributions at $\sqrt{s}=630$
GeV obtained by P238 \cite{P238} (open circles) and UA7 \cite{UA7}
collaborations
(full circles) with predictions of our model.}
\end{figure}
\vspace{-1cm}\\

It is well known that the charged particle multiplicity distributions
may well be fitted by a negative binomial distribution (NB) \cite{Gim}.
It has two parameters: mean charged multiplicity $\bar{n}$, and the
parameter $k$ $(k\geq 1)$ affecting its shape (width) (for $k
\rightarrow 1$ NB approaches geometrical distributions whereas
for $k^{-1}\rightarrow 0$ it approaches Poisson distribution). The
observed widening of the normalized multiplicity distribution with
increasing energy implies then that parameter $k$ decrreases with
energy.
Following ideas expressed in \cite{WWq} we would like to bring ones
attention to the fact that the value of parameter $k^{-1}$ may be
understood as the measure of fluctuations of mean multiplicity. When
one starts with Poisson multiplicity distribution and then allows
$\bar{n}$ to fluctuate according to gamma distribution with
normalized variance given by $D(\bar{n})$ then it is easy to show
\cite{WWq} that as result one gets NB distribution with $k^{-1} =
D(\bar{n}) = \sigma^2(\bar{n})/\langle \bar{n}\rangle^2 = q - 1$.
Namely:\\
\vspace{-3.4cm}
\begin{figure}[h]
\setlength{\unitlength}{1cm}
\begin{picture}(8.0,8.0)
\includegraphics{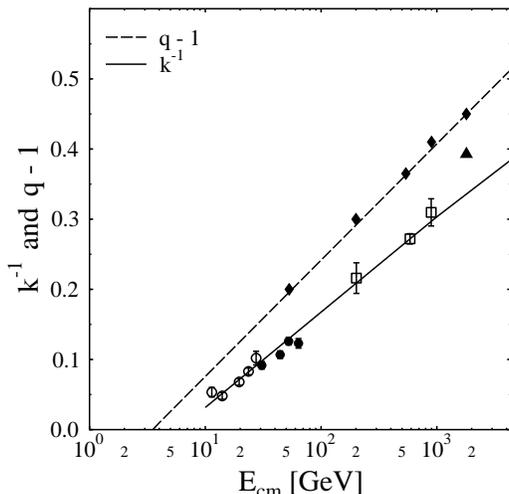}
\end{picture}
\caption{The full line shows the energy dependence of $k^{-1}$
obtained from NB fits to multiplicity distributions at collider
energies \cite{Gim}. The dashed line shows the energy variation of
our parameter $q$ obtained from our fits in Fig. 1.}
\end{figure}
\vspace{-1cm}\\
\begin{eqnarray}
P\left( n \right) & = &
\int_{0}^{\infty}\frac{e^{-\overline{n}}\cdot
\overline{n}^{n}}{n!} \cdot \frac{\gamma ^{k} \cdot
\overline{n}^{k-1} \cdot e^{-\gamma
\overline{n}}}{\Gamma\left(k\right)}d\overline{n}=\nonumber\\
& = & \frac{\Gamma \left(k+n \right)}{\Gamma \left(1+n \right)
\Gamma \left(k \right)} \cdot \frac{\gamma^{k}}{\left(\gamma + 1
\right)^{k+n}}, \label{eq:PNB}
\end{eqnarray}
where $\gamma = k/\langle \overline{n}\rangle   $. It
is worth to notice that, indeed, the energy variation of parameter
$q$ is almost the same as $k^{-1}$ (see Fig. 3). The small
discrepancies observed there, namely the fact that $q-1$ is
consistently larger than $k^{-1}$, can be explained by realizing that
our $q$ has been obtained from fitting the experimental distributions
$dN/dy$ and therefore it contains in addition to the above
mentioned fluctuations in $\bar{n}$ also fluctuations in inelasticity
$K$, not discussed here. In any case it is at this moment
tempting to assume that the nonextensive parameter $q$ (parametrizing
already nonlinearity of our FP equation used to describe data) also
describes those fluctuations.

In Fig. 4 we show the total inelasticity $K = K(s)$ obtained by
integrating spectra given by eq. (\ref{eq:Fq}) with parameters
obtained from the fit to the corresponding rapidity distributions
\begin{equation}
K = \frac{2}{\sqrt{s}}\, \int^{Y_{max}}_0\, \frac{3}{2}\cdot
f_q(y)\cdot m_T \cosh y\, dy \label{eq:inel}
\end{equation}
as function of the invariant energy of the reaction $\sqrt{s}$ both
for the energies considered here and extrapolated to the LHC energy.
In this case the values of transverse mass $m_T$ for a given energy
was obtained by using simple interpolating formula: $m_T = 0.3 +
0.044 \ln(\sqrt{s}/20)$. As one can see in Fig. 4 such inelasticity
$K(s)$ is decreasing with the energy. \\
\vspace{-3.5cm}
\begin{figure}[h]
\setlength{\unitlength}{1cm}
\begin{picture}(8.0,8.0)
\includegraphics{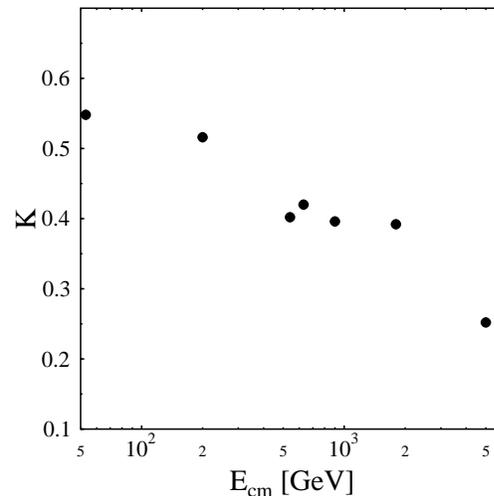}
\end{picture}
\caption{The energy dependence of total inelasticity obtained from
our model.}
\end{figure}
\vspace{-0.8cm}

\section{SUMMARY AND CONCLUSIONS}

We have provided here description of rapidity spectra of particles
produced at CERN and Fermilab energies treating their formation as
diffusion process in rapidity space. To account for the anomalous
character of such diffusion, the nonlinear form of Focker-Planck
equation has been used with nonlinearity described by parameter $q$,
the same as the nonextensivity parameter describing the underlying
Tsallis statistics. As seen in Figs. 1 and 2, very good agreement
with data has been obtained with apparently three parameters: the
"temperature" $T$, position of the peak at rapidity $y_m$ and
parameter $q$ - all logaritmically dependent on the energy of
reaction $\sqrt{s}$ \footnote{We have checked that we could not get
such good fits within formalism with $q=1$, what means that
nonlinearity expressed by $q>1$ here is essential feature of the
data.}. However, after closer inspection it turns out that parameter
$q$, which according to \cite{WWq} can be regarded as a measure of
fluctuations existing in the physical system under consideration,
follows essentiall the fluctuations of multiplicity of particles
produced at given energy. The small differences noticed in Fig. 3
are, in our opinion, caused entirely by the fluctuations in the
inelasticity of the reaction, which makes the initial energy
available for the production of secondaries a fluctuating
quantity\footnote{However, we did not attempt here to estimate such
fluctuations. We are planning to do it elesewhere.}. Similarly,
parameter $y_m$ seems to be closely connected with the maximal
available rapidity $Y_{max}(\sqrt{s})$ \footnote{There is therefore
possibility that it is, in fact, connected with $Y_{max}(K\cdot
\sqrt{s})$, i.e., it depends in the indirect way on the inelasticty
$K$ of the reaction. We have not pursued this problem here.}. This
shows that the only parameter which is entirely "free" is the
limiting temperature $T$ \footnote{One should notice at this point
that its energy dependence as given in eq.(\ref{eq:param}) contains
in itself also the possible energy dependence of the mean transverse
mass $m_T$, not accounted for in the formula (\ref{eq:Fq}).}.
Therefore we can say that what we are proposing here is essentially
one-parameter fit successfully describing data on rapidity
distributions.

The resultant inelasticity $K$ (defined by the formula
(\ref{eq:inel})) turns out to be decreasing function of energy, as
seen in Fig. 4, confirming our previous findings in that matter
\cite{inel}. We would like to bring ones attention to the fact that
in our approach we can fit equally well (using the same set of
parameters) data obtained by P238 \cite{P238} and UA7 \cite{UA7}
collaborations which cover different regions in rapidity. This makes
extrapolation of our formula to higher rapidity region more credible
and therefore puts more weight on the obtained energy behaviour of
inelasticity.

Acknowledgements: The partial support of Polish Committee for
Scientific Research (grants 2P03B 011 18 and
621/E-78/SPUB/CERN/P-03/DZ4/99) is acknowledged.\\

\end{document}